\documentclass[sigconf, authorversion, natbib=false]{acmart}
\AtBeginDocument{%
  }

\setcopyright{acmlicensed}

\copyrightyear{2026}
\acmYear{2026}
\setcopyright{cc}
\setcctype{by}
\acmConference[SAC '26]{The 41st ACM/SIGAPP Symposium on Applied Computing}{March 23--27, 2026}{Thessaloniki, Greece}
\acmBooktitle{The 41st ACM/SIGAPP Symposium on Applied Computing (SAC '26), March 23--27, 2026, Thessaloniki, Greece}
\acmPrice{}
\acmDOI{10.1145/3748522.3779786}
\acmISBN{979-8-4007-2294-3/2026/03}

\RequirePackage[
  datamodel=acmdatamodel,
  style=acmnumeric,
  ]{biblatex}

\usepackage{float}
\usepackage{booktabs}
\usepackage{caption}
\usepackage{amsmath,amsfonts}
\usepackage{algorithmic}
\usepackage{graphicx}
\usepackage{textcomp}
\usepackage{xcolor}
\usepackage{algorithm}
\usepackage{amsthm}
\usepackage{url}

\DeclareMathOperator{\argmin}{\arg\!\min}

\begin{document}


\title{Linear Complexity Self-Supervised Learning for Music Understanding with Random Quantizer}



\author{Petros Vavaroutsos}
\orcid{0000-0003-1929-5649}
\email{peter@orfium.com}
\affiliation{%
\institution{Orfium Research}
  \city{Athens}
  \state{}
  \country{Greece}
}

\author{Theodoros Palamas}
\orcid{0009-0001-5846-839X}
\email{theodoros@orfium.com}
\affiliation{%
  \institution{Orfium Research}
   \city{Athens}
  \state{}
  \country{Greece}
}

\author{Pantelis Vikatos}
\orcid{0000-0003-1573-5125}
\email{pantelis@orfium.com}
\affiliation{%
  \institution{Orfium Research}
  \city{Athens}
  \state{}
  \country{Greece}
}

 \renewcommand{\shortauthors}{}
 \renewcommand{\shortauthors}{P. Vavaroutsos et al.}
 
\begin{abstract}
{In recent years, foundation models have become very popular due to their exceptional performance, mainly in natural language (NLP) tasks where they were first introduced. These models usually consist of hundreds of millions, or even billions, of parameters, making them resource-intensive during training and in production systems, leading to increased costs. This paper focuses on the reduction of a foundation's model size when applied to music information retrieval (MIR) tasks. Our research combines the Branchformer architecture with SummaryMixing, which were first applied in speech recognition, along with a random quantization process. To facilitate reproducibility, we conduct pre-training on publicly available datasets, complemented by a proprietary dataset comparable in scale to other private datasets reported in the literature. We ensure robust evaluation by using a framework consisting of a variety of downstream MIR tasks. Our results show that our architecture achieves competitive performance when compared with other state-of-the-art models that use multi-head self-attention, while reducing the model size from $8.5$\% up to $12.3$\%.}

\end{abstract}

\ccsdesc[500]{Computing methodologies~Machine learning approaches}






\keywords{Deep Learning, Learnable Representations, Music Understanding, Transformers, Embeddings, Attention}

\maketitle

\section{Introduction}\label{sec:introduction}
Foundation models have become a core element in deep learning research due to their exceptional performance across a wide range of tasks. They initially emerged from the field of natural language processing (NLP) with the GPT~\cite{radford2018improving,radford2019language,brown2020language} and BERT~\cite{devlin2018bert} architectures and were soon applied to the field of 
speech recognition~\cite{gulati2020conformer,peng2022branchformer} and music~\cite{li2023mert}. The power of a foundation model comes from its pre-training on large corpuses of data using a self-supervised manner with the aim of discovering underlying structures in the data and learning generalizable vector representations. The knowledge accumulated in pre-training can later be leveraged in a wide range of downstream tasks. This usually involves a frozen pre-trained foundation model extended with an extra layer trained for a more specific task.

Foundation models in the field of music information retrieval are in their early stages, with MERT~\cite{li2023mert} currently being at the forefront. It is an early effort to tackle the large variety of MIR tasks in a general manner rather than individually. It uses self-supervised learning with masked language modelling to train a transformer model based on HuBERT~\cite{hsu2021hubert} on a large audio dataset. In the pre-training stage they use two teachers to guide the model. The first is an RVQ-VAE~\cite{zeghidour2021soundstream, defossez2022high} which is used to extract quantized audio token representations and the second is the ConstantQ-Transform (CQT) spectral representation of the audio. In the pre-training stage the foundation model tries to learn both the quantized representations and reconstruct the CQT. They show that a general foundation model can achieve state-of-the-art performance on MIR tasks when compared to models trained specifically for each task.

Despite the significant success of foundation models across various domains, they continue to pose several challenges, with one of the most prominent being their sheer size. The 
large
scale of these models, while contributing to their high performance, also introduces substantial computational and resource requirements, both during the training phase and at deployment. 
As a result, optimizing their size without sacrificing performance has become a key research focus. 
Efforts to reduce the size and the training time of foundation models are being explored across multiple domains, including natural language processing, computer vision, and more. 
In our work, we specifically address this challenge in the domain of music. 
We adopt well known transformer backbones, from speech domain, paired with linear complexity self-attention alternatives and random quantization process, forming a new architecture, specifically tailored for the music domain.
Each of those choices has been shown to reduce either the size, the training time or the memory footprint of a foundation model in their respective works, with good performance in speech recognition tasks. 
To the best of our knowledge this is the first research effort that applies the aforementioned architecture in MIR. Our code and models are made freely available at https://github.com/Orfium/muse-lq


We structure the rest of the paper as follows. Section~\ref{sec:Related Work} includes the related literature. Section~\ref{sec:Model Overview} contains the model overview, presenting the details of the model layers. Section~\ref{sec: Experiments} contains a reference to our experimentation setup: the datasets, the  description of the downstream tasks as well as details of the training hyperparameters. In section~\ref{sec:Results}, we provide an overview of the results and contributions. Finally, in section~\ref{sec:discussion} we discuss about the limitations of this work and future work, and section~\ref{sec:conclusions} concludes with the highlights and an outlook on future work.

\section{Related Work}\label{sec:Related Work}

Recent advancements in foundation models, such as GPT, BERT, and their variants, have demonstrated remarkable capabilities across a wide range of domains, including natural language processing (NLP), computer vision, and music information retrieval (MIR). These models, with their ability to learn and generalize from vast amounts of data, have revolutionized fields like text generation, language understanding, image recognition, and more. In the domain of MIR, which deals with the extraction, analysis, and understanding of information from music, the application of foundation models is still in its early stages. However, initial research in this area has proven to be highly promising, suggesting that these models have the potential to significantly improve the way we process and interact with music data.

\subsection{CNNs and Contrastive Learning}

Early deep-learning attempts have mainly focused on investigating the performance of supervised CNN architectures when applied to the domain of MIR. The research effort~\cite{pons2016experimenting} experiments with the filter size of CNNs in both dimensions, using mel-spectrograms as input, in order to capture both the temporal and frequency relationships present in the audio.
Study~\cite{pons2017end} investigates the effect that different input features have on performance by comparing raw waveforms with mel-spectrograms. The aforementioned models are compared in~\cite{pons2019musicnn} along with VGG-like models~\cite{simonyan2014very} in music tagging tasks.

Apart from supervised learning, unsupervised contrastive methodologies have also been used, originating from the field of vision~\cite{chopra2005learning, schroff2015facenet, chen2020simple}. These methods focus on learning well-structured embedding representations of the data instead of focusing on a specific supervised task. SimCLR~\cite{chen2020simple} was later employed in MIR~\cite{spijkervet2021contrastive} for learning meaningful musical representations. For each audio waveform, the model generates an augmented and correlated version by applying a series of transformations, such as adding noise, pitch-shifting, or frequency filtering. The augmented waveform is then used as the positive sample in a contrastive loss framework, where it is paired with the original, along with an unrelated audio from the same batch as a negative sample. This approach encourages the model to learn similar representations for both the original and augmented versions. 
A comparative analysis between several supervised and unsupervised strategies is provided in study~\cite{mccallum2022supervised}.

They use a variety of datasets, representing a wide range of MIR tasks. They conclude that unsupervised learning produces models that generalize efficiently in more tasks than supervised learning, which excels when using annotated music data from experts. Furthermore, they show that the dataset used for pre-training has a significant impact on performance, especially when it restricts the audio to a specific domain.

\subsection{Quantization}

Additionally, there is a focus on investigating discrete tokenized representations in comparison with continuous embedding representations. Among those is VQ-VAE~\cite{van2017neural}, a new type of variational autoencoder which incorporates vector quantization. Their study shows that tokenized representations can model long-term relationships and have similar performance to continuous representations. Jukebox~\cite{dhariwal2020jukebox} combines multiple VQ-VAEs trained on different temporal resolutions, along with a Transformer architecture for the generation of music. To avoid codebook collapse, which affects VQ-VAEs, they additionally use random resets on codebook words during training.
The discrete representations learnt by Jukebox improve performance when used in a variety of MIR tasks, namely JukeMIR~\cite{castellon2021codified}.

MERT~\cite{li2023mert} uses a HuBERT-based architecture~\cite{hsu2021hubert} which is trained using an RVQ-VAE~\cite{zeghidour2021soundstream, defossez2022high} for tokenization and a CQT reconstruction loss. HuBERT was first introduced in the field of speech recognition as a way to apply BERT to speech, by extending it with a convolutional encoder and using a k-means clustering on the audio MFCCs as training targets. The RVQ-VAE is an extension of the VQ-VAE, that calculates the residuals after the initial quantization and then uses an additional codebook to quantize them. This process is then repeated multiple times. Study~\cite{won2024foundation} further evaluates MERT along with the Conformer architecture~\cite{gulati2020conformer} by using a random projection quantizer instead~\cite{chiu2022self} which removes the need for the training of the quantizer while maintaining state-of-the-art performance.
OMAR-RQ model~\cite{alonso2025omar} extended the approach of Won et al.~\cite{won2024foundation} by adopting a multilabel setting with parallel residual quantization codebooks, enabling  music audio representations. Won et al.~\cite{zhu2025muq} proposed MuQ, a self-supervised music representation model, that uses Mel Residual Vector Quantization (Mel-RVQ) to learn compact and expressive audio features.

\subsection{Overlap with Speech Recognition}

Due to the similarities between the speech recognition and MIR fields~\cite{jasmin2020tailored}, methodologies from the latter are already being applied on the former. For example, AudioLM~\cite{borsos2023audiolm} is shown to be capable of generating both speech and music. In order to achieve this, the authors of this work use both semantic and acoustic tokenization methods from literature in a hierarchical process, with each level of tokenization assisting in the production of the next one. They show that AudioLM is not only capable of generating coherent speech, but also effective in producing piano continuations of high quality. Conformer~\cite{gulati2020conformer} and Branchformer~\cite{peng2022branchformer} combine the ability of Transformers to model long-term context relationships with the ability of CNNs to capture local features, achieving state-of-the-art performance on speech recognition tasks. The Conformer and Branchformer architectures consist of an initial convolutional subsampling component, used for input dimensionality reduction, followed by a series of Transformer-inspired blocks. The Conformer blocks are based on the Macaron architecture~\cite{lu2019understanding} with the addition of a convolutional layer, combining multi-head self-attention and convolution linearly. The Branchformer blocks, on the other hand, combine them in parallel. These blocks use a multi-head self-attention branch and a convolutional branch~\cite{sakuma2022mlp}, which merge at the end. Other works investigate reducing inference time, by using linear complexity alternatives to self-attention like SummaryMixing~\cite{parcolletsummarymixing}, Fastformer~\cite{wu2021fastformer}, HyperMixing~\cite{mai2023hyperconformer} and Mamba~\cite{gu2023mamba}, and training time by pairing those with random quantization methods~\cite{whetten2024analysis}.

Although foundation models have found application in many domains, efforts to explore their viability in MIR tasks are still in their infancy. With our work, we aim to expand their application by investigating potential ways to reduce their size, along with their training and inference time, while maintaining performance. We investigate both Conformer and Branchformer backbones~\cite{gulati2020conformer,peng2022branchformer}, which were initially applied in speech recognition and achieved state-of-the-art results. We expand our research with a linear complexity module alternative to multi-headed self-attention in transformers~\cite{parcolletsummarymixing}, which has shown that it can reduce both the inference time and memory footprint of the model, while maintaining performance, in the speech domain. We also use a non-trainable random tokenization process~\cite{chiu2022self}, significantly reducing the training time. Finally, we compare our work against other state-of-the-art works such as MERT and other studies, where the Conformer architecture was evaluated but only on a narrow set of tasks, such as beat tracking, chord recognition, key detection and more~\cite{won2024foundation}. In our work, we expand this evaluation in an extended MIR downstream task suite.

\section{Model}\label{sec:Model Overview}

\begin{figure}[t!]
    \centering
    \includegraphics[width=8cm]{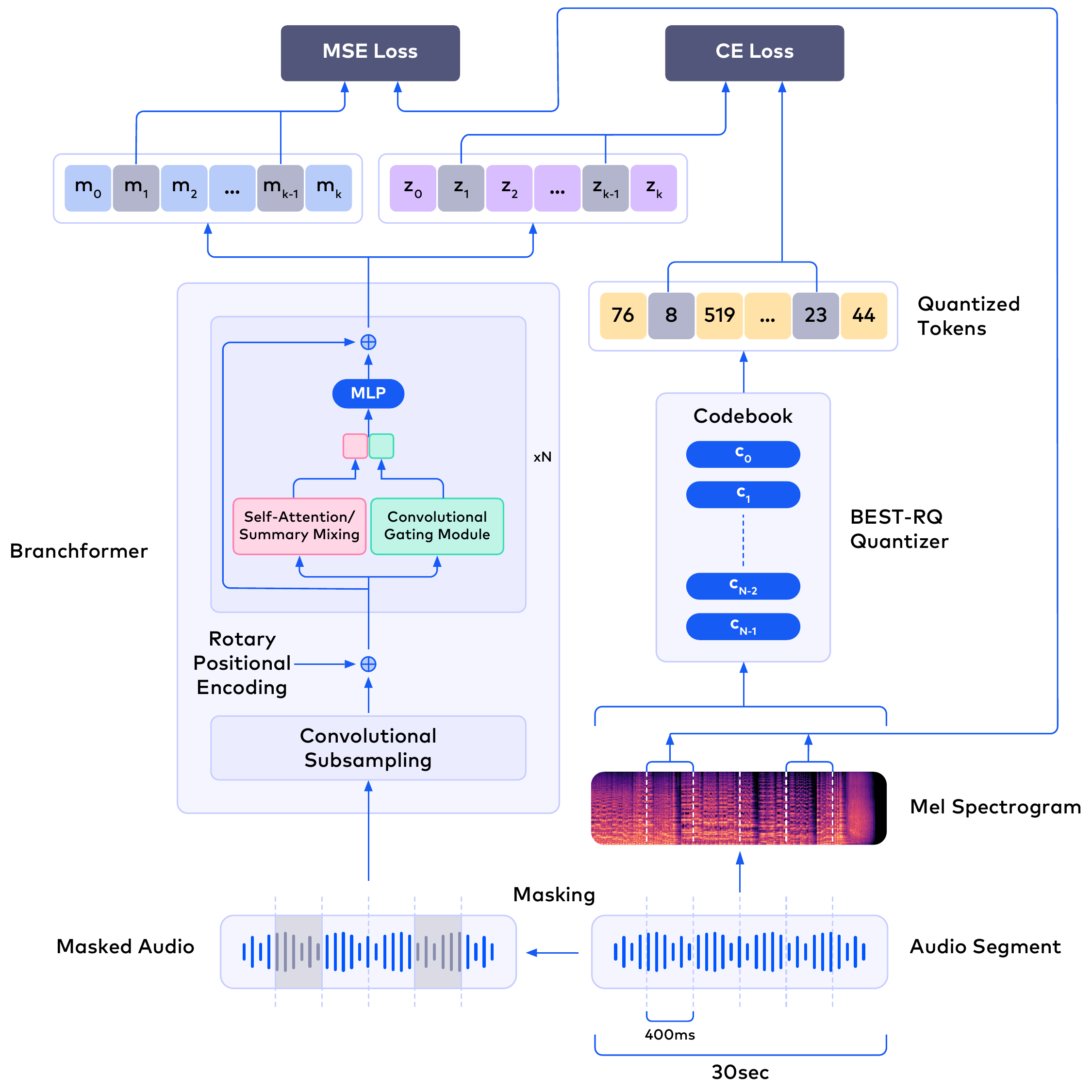}
    \Description{Model Overview}
    \caption{An illustration of our framework is shown here. On the left, we present the Branchformer architecture, which takes raw masked audio as input and generates two distinct sets of logits. Specifically, $m_0, ..., m_{k}$ represent the logits corresponding to the mel spectrogram, while $z_0, ..., z_{k}$ correspond to the logits for the quantized tokens. On the right, we show the random tokenizer, which receives the mel spectrogram as input and produces the quantized tokens. Importantly, only the masked tokens are utilized for the loss calculation, ensuring that the model focuses on the relevant parts of the input.}
    \label{fig:pipeline}
\end{figure}

\subsection{Architecture}
We focus on a Branchformer architecture coupled with SummaryMixing as seen in Figure~\ref{fig:pipeline}. It consists of a convolutional subsampling component followed by $N$ Branchformer encoder blocks. The convolutional component is used to downsample the waveform, after which we also add rotary positional encodings~\cite{su2024roformer}, when using multi head attention. The encoder blocks consist of two branches, one using multi-head self-attention in order to capture global dependencies and one using a convolutional gating module~\cite{sakuma2022mlp} which captures local dependencies. Finally, the two branches are concatenated due to better performance on a weighted average~\cite{peng2022branchformer}. Capturing both global and local dependencies is beneficial compared to an attention-only module~\cite{gulati2020conformer,peng2022branchformer}.

\subsection{Linear Complexity Alternative}
From the conception of the Transformer, the self-attention module has been a key part of its structure. While these modules are the reason for the exceptional performance of Transformers, they still have some setbacks. The computations of multi-head self-attention modules are quadratic in complexity with respect to the sequence length. Efforts are already being made to find alternatives with reduced complexity, with one of them being SummaryMixing~\cite{parcolletsummarymixing}. They show in their work that by substituting the self-attention module with a SummaryMixing module, the computation complexity drops down to linear, making training faster and reducing memory footprint. They experiment with substituting the multi-head self-attention modules of both Conformer and Branchformer models with their proposed solution. We similarly adopt SummaryMixing in the scope of MIR and compare it with the standard self-attention module.

\subsection{Tokenization}
Tokenization is the process by which the model input data are split into smaller discrete segments in order to simplify their structure and dimensionality. Methodologies for creating discrete audio representations from continuous data has already been the subject of study in literature. HuBERT applies a k-means clustering on the MFCCs, extracted from the audio, and uses the cluster centroids for discretization, while vq-wav2vec~\cite{baevski2019vq} extends wav2vec~\cite{schneider2019wav2vec} representations for a similar purpose. The drawback of these methods is that they need to be trained. This increases computation costs and biases the tokenizer towards the train dataset distribution.

The tokenization method consists of two components: a randomly generated codebook and a randomly generated projection to the codebook words~\cite{chiu2022self}. Thus, the tokenizer is not trainable, which alleviates the learning procedure by being constant after its initialization. Another benefit of this process is that it does not suffer from codebook collapse, a state in which all codebook words converge into a single or few vectors. The tokenization process is as follows:
\[y=\argmin_{i}||norm_{l2}(c_i)-norm_{l2}(A \cdot x)||\]
where $y$ is the discrete token, $x$ is our $d$-dimensional input vector, $A$ is the randomly initialized $h \times d$ projection matrix and $c_i$ is one of $n$ words from the randomly initialize codebook $C=\{c_0,...,c_{n-1}\}$. The projection matrix $A$ uses Xavier initialization~\cite{glorot2010understanding} and the
codebook $C$ uses standard normal distribution for initialization.

\subsection{Masking}
In the context of training deep learning models, masking involves any process by which parts of the input data are obscured. The model is then tasked with predicting the hidden information. This process makes the model robust to incomplete or noisy data and pushes it to learn contextual relationships.

When dealing with spectral representations of audios, the most common masking techniques are frequency masking, temporal masking, or a combination of both. Frequency masking involves setting to zero certain frequency bands, which forces the model to make predictions based on a frequency context. On the other hand, time masking involves setting to zero audio segments based on time. Thus, the model tends to learn temporal patterns. These can also be combined to enforce both frequency and temporal contexts. Furthermore, they can be used in conjunction with other audio modifications, as shown in SpecAugment~\cite{park2019specaugment}, where masking is used in conjunction with time warping.

In our case, due to the tokenized nature of our data, we apply only random time masking to the waveform input as shown in Figure \ref{fig:pipeline}. We split the waveform in $400ms$ segments, which are then randomly chosen with a probability of $20\%$. The discrete tokens corresponding to those chosen segments are subsequently masked. The masking for each input audio changes in every epoch.

\subsection{Loss}
For our self-supervised pre-training scheme, we use two training objectives: a reconstruction loss and a classification loss. The purpose of the reconstruction loss is for the model to capture the low-level audio features by trying to predict the original mel-spectrogram, potentially by constructing an internal compact representation. On the other hand, the classification loss pushes the model to learn the quantized tokens of the audio segments, as specified by the codebook. This representation is discrete and high-level, in contrast with the mel-spectrogram features. The combination of these objectives makes the model more robust and generalizable when used in downstream tasks later. We use a mean squared error loss ($MSE$) for reconstruction and a cross entropy loss ($CE$) for classification.

Assuming $l$ are the model logits and $t$ are the target tokens generated from our tokenizer, the loss is computed as follows:
\[L=CE(softmax(l_t),t_t)+MSE(l_{mel},t_{mel})\]
We only calculate the loss for the masked tokens. The purpose of this is to focus the model on learning only from missing information. In addition, it requires less computations, making training efficient. Using all the tokens has the potential to dilute the effect of masking and potentially make the model prone to overfitting.

\section{Experiments}
\label{sec: Experiments}
\subsection{Music Datasets}

For training, we utilize three distinct datasets, including one private dataset and two publicly available ones. The private dataset offers the advantage of being curated to ensure higher quality, and it contains a significantly larger amount of music samples, spanning a wide range of durations. This allows for more comprehensive and diverse training. On the other hand, using public datasets enables us to benchmark our results against those of other works in the literature, fostering comparability and ensuring that our findings are consistent with existing research.

Music4All~\cite{santana2020music4all} is a publicly available dataset that contains around $910$ hours of audio collected from online platforms (YouTube / Spotify / Last.fm) and was sanitized in order to be of high quality. It contains $30$ second audio segments and their corresponding metadata, such as artist, popularity, energy, tempo and others.

The other public dataset is FMA~\cite{defferrard2016fma} which is a dump of the free and open library Free Music Archive\footnote{https://freemusicarchive.org/} and contains audio alongside low \& high-level features. It has several subsets with different sizes. In our work, we use the fma-large subset which contains around $900$ hours of audio.

Lastly, we gather and curate a high quality proprietary dataset of around $200$k hours of music. Its size ensures that it is comparable to other private datasets mentioned in the literature, such as the one in MERT which is around $160$k hours. Table \ref{table:datasets} shows a summary of the datasets.

\begin{table}
\resizebox{\columnwidth}{!}{%
\begin{tabular}{|l|r|r|}
\hline
\textbf{Dataset} & \multicolumn{1}{l|}{\textbf{Hours}} & \multicolumn{1}{l|}{\textbf{Audio Duration}} \\ \hline
Music4All 
& $910$ & $30$sec \\
FMA-large 
& $900$ & $30$sec \\
Proprietary & $\sim200,000$ & $\sim3-4$min (full songs) \\
MERT Proprietary 
& $\sim160,000$ & $\sim3-4$min (full songs) \\ \hline
\end{tabular}%
}
\caption{Details for the datasets used in our experiments. Audios in our proprietary dataset are split into $30$sec segments without overlap to match the duration of public datasets.}
\label{table:datasets}
\end{table}

\subsection{Evaluation}

\subsubsection{Model Comparison}
Our proposed architecture consists of the Branchformer model in which the multi-head self-attention module is replaced with SummaryMixing. We compare this model against a vanilla Branchformer. 
Also, we evaluate two versions of a Conformer model~\cite{gulati2020conformer, won2024foundation}, one with multi-head self-attention, i.e. vanilla, and one with SummaryMixing. Lastly, we benchmark a transformer-based model i.e. MERT-330~\cite{li2023mert}.

\subsubsection{Downstream Tasks}

The models are evaluated in a variety of classification and regression MIR downstream tasks which are used in the previous studies. 
After pre-training our foundation models, we freeze their parameters and attach at the end one dense layer with dropout. This layer is then trained for each of the following downstream tasks. In the following tasks, wherever not specified, the dataset split is $80$\% (train) / $10$\% (validation) / $10$\% (test).

\textbf{Music Tagging} consists of the prediction of various high-level audio features such as genre, mood, instrument and others. While these tag subcategories can be evaluated separately, their joint use gives us an insight about a model's ability to be robust while trying to predict all of them at the same time. We use two common public datasets, MagnaTagAtune~\cite{law2009evaluation} and MTG-Jamanedo~\cite{bogdanov2019mtg}. We use a window length of $30$ seconds and ROC-AUC and Average Precision (AP) as metrics.

\textbf{Key Detection} is the prediction of the tonal scale and pitch of a song. The classes are the $12$ major keys, $12$ minor keys and None. We use the  Giantsteps-MTG-keys dataset~\cite{korzeniowski2017end} for downstream training and Giantsteps~\cite{knees2015two} for testing. The metric is refined accuracy as implemented in mir\_eval~\cite{raffel2014mir_eval}, which is tolerant to reasonable prediction errors.

\textbf{Genre Classification} aims to predict the most likely genre of a song. We evaluate on GTZAN~\cite{tzanetakis2002musical} and the genre subset of MTG-Jamendo~\cite{bogdanov2019mtg}. For GTZAN we use the fail-filtered split~\cite{kereliuk2015deep} and accuracy as metric. For MTG-Jamendo we use ROC-AUC and AP.

\textbf{Emotion Score Regression} is a task where we want to predict the valence (positive/negative emotional response) and arousal (emotional intensity) of a song. The Emomusic dataset~\cite{soleymani20131000}   contains $744$ $45$-second audios where humans reported the valence and arousal after listening to them. The official metric used is determination coefficient ($r^2$) between the model predictions and the actual human responses.

\textbf{Instrument Classification} is another important audio tagging task, where the goal is to identify the specific instruments present in a given audio sample. This task involves analyzing the audio's features to determine which instruments are playing, whether individually or in combination. We use NSynth~\cite{engel2017neural} containing $11$ instruments where we report the accuracy. We also use the instrument subset of MTG-Jamendo~\cite{bogdanov2019mtg} which contains $41$ instruments, where we report the ROC-AUC and AP.

\textbf{Pitch Classification} is the process of estimating the pitch class of a given audio sample, determining which of the $128$ distinct pitch categories the sample belongs to. This task involves analyzing the audio's features to make an accurate prediction. For our evaluation, we use the NSynth dataset~\cite{engel2017neural} and report the resulting accuracy to assess the performance of the model.

\textbf{Vocal Technique Detection} aims to identify the vocal technique used by the singer in a song. We use Vocalset dataset~\cite{wilkins2018vocalset} which contains the vocals of $20$ different professional singers who perform $17$ different vocal techniques. We used the same vocal techniques and train/test splits as in~\cite{yamamoto2022deformable}. However, no validation split is provided. For this reason we set aside for validation $10$\% of the training set. The evaluation metric is accuracy.

\textbf{Singer Identification} involves predicting the performer of a given audio track based on its characteristics. For this task, we utilize the Vocalset dataset~\cite{wilkins2018vocalset} once again and present the resulting accuracy as part of the evaluation.

\subsection{Experimental Setup}

\begin{table}
\resizebox{\columnwidth}{!}{%
\begin{tabular}{|l|r|l|l|}
\hline
\textbf{Model} & \multicolumn{1}{l|}{\textbf{Parameters}} & \textbf{\begin{tabular}[c]{@{}l@{}}Encoder\\ Layers\end{tabular}} & \textbf{\begin{tabular}[c]{@{}l@{}}Encoder\\ Dimension\end{tabular}} \\ \hline
CNF$^{\triangle att}$ & $95m$ & $4$ & $768$ \\
CNF$^{\blacktriangle att}$ & $330m$ & $12$ & $1024$ \\
CNF$^{\blacktriangle sum}$ & $302m$ & $12$ & $1024$ \\
BRN$^{\triangle att}$ & $60m$ & $4$ & $768$ \\
BRN$^{\blacktriangle att}$ & $187m$ & $12$ & $1024$ \\
BRN$^{\blacktriangle sum}$ & $164m$ & $12$ & $1024$ \\ \hline
\end{tabular}%
}
\caption{Comparison of different model configurations. $\triangle$ denotes a small model while $\blacktriangle$ denotes a large model. $att$ and $sum$ are the attention and SummaryMixing versions respectively.}
\label{table:models}
\end{table}
The model hyperparameters used in our experiments are detailed in Table \ref{table:models}. Our models were trained across a distributed setup of $64$ A100 GPUs, each with $40$ GB of memory, utilizing bf16 precision for improved performance and memory efficiency. Due to memory constraints, we opted for a local batch size of $16$ samples per GPU, which results in a global batch size of $1024$ across all GPUs. Each batch represents approximately $8.5$ hours of audio. We used the Adam optimizer, starting with a linear warmup phase that gradually increased the learning rate up to $10^{-4}$, followed by cosine decay, where the learning rate is reduced to $10^{-5}$. During the early stages of training, particularly when training our larger models, we observed issues with exploding gradients after a certain number of steps. To mitigate this, we implemented gradient clipping with a threshold of $1.0$, which helped improve the training process and prevent training instability.

Our audio files are divided into $30$-second segments, each with a sample rate of $24000$ Hz. During preprocessing, we randomly mask $400$-millisecond windows with a probability of $20\%$. To extract the mel-spectrograms, we use a Fast Fourier Transform (FFT) with a window size of $2048$, a hop length of $240$ samples, and the default $128$ mel bins. This setup produces a mel-spectrogram with dimensions of $128 \times 100$, corresponding to $100$Hz frequency samples. To reduce the temporal resolution, we stack intermediate time points with a factor of $4$, resulting in a mel-spectrogram of size $512 \times 25$, which corresponds to $25$Hz temporal samples. For tokenization, we employ the random tokenizer, using a codebook size of $8192$ and a codebook dimension of $16$.

Finally, for downstream training, an additional dense layer is included, which contains $512$ units and a dropout probability of $0.25$ to help prevent overfitting. The batch size is determined based on the specific task, with values set to either $16$ or $32$ to optimize performance and training efficiency.
\section{Results}\label{sec:Results}

\begin{table*}[]
\resizebox{\textwidth}{!}{%
\renewcommand{\arraystretch}{1.1}
\begin{tabular}{|l|llll|l|lll|ll|lll|l|l|l|}
\hline
\multicolumn{1}{|c|}{\textbf{Task}}
 & \multicolumn{4}{c|}{\textbf{\begin{tabular}[c]{@{}l@{}}Music Tagging\end{tabular}}} & \textbf{\begin{tabular}[c]{@{}c@{}}Key\\ Detection\end{tabular}} & \multicolumn{3}{c|}{\textbf{\begin{tabular}[c]{@{}c@{}}Genre\\ Classification\end{tabular}}} & \multicolumn{2}{c|}{\textbf{\begin{tabular}[c]{@{}c@{}}Emotion\\ Score\\ Regression\end{tabular}}} & \multicolumn{3}{c|}{\textbf{\begin{tabular}[c]{@{}c@{}}Instrument\\ Classification\end{tabular}}} & \textbf{\begin{tabular}[c]{@{}c@{}}Pitch\\ Classification\end{tabular}} & \textbf{\begin{tabular}[c]{@{}c@{}}Vocal\\ Technique\\ Detection\end{tabular}} & \textbf{\begin{tabular}[c]{@{}c@{}}Singer\\ Identification\end{tabular}} \\ \cline{1-17} 
\multicolumn{1}{|c|}{\textbf{Dataset}} & \multicolumn{2}{c|}{\textbf{MTAT}} & \multicolumn{2}{c|}{\textbf{MTG}} & \multicolumn{1}{c|}{\textbf{Giantsteps}} & \multicolumn{1}{c|}{\textbf{GTZAN}} & \multicolumn{2}{c|}{\textbf{MTG}} & \multicolumn{2}{c|}{\textbf{Emomusic}} & \multicolumn{1}{c|}{\textbf{NSynth}} & \multicolumn{2}{c|}{\textbf{MTG}} & \multicolumn{1}{c|}{\textbf{NSynth}} & \multicolumn{1}{c|}{\textbf{Vocalset}} & \multicolumn{1}{c|}{\textbf{Vocalset}} \\ \cline{1-17} 
\multicolumn{1}{|c|}{\textbf{Metric}}
 & \multicolumn{1}{c|}{ROC} & \multicolumn{1}{c|}{AP} & \multicolumn{1}{c|}{ROC} & \multicolumn{1}{c|}{AP} & \multicolumn{1}{c|}{Acc} & \multicolumn{1}{c|}{Acc} & \multicolumn{1}{c|}{ROC} & \multicolumn{1}{c|}{AP} & \multicolumn{1}{c|}{$r^2_A$} & \multicolumn{1}{c|}{$r^2_V$} & \multicolumn{1}{c|}{Acc} & \multicolumn{1}{c|}{ROC} & \multicolumn{1}{c|}{AP} & \multicolumn{1}{c|}{Acc} & \multicolumn{1}{c|}{Acc} & \multicolumn{1}{c|}{Acc} \\ \hline
\hline
CNF$^{\blacktriangle att}_{fma}$ & \multicolumn{1}{c|}{0.879} & \multicolumn{1}{c|}{0.421} & \multicolumn{1}{c|}{0.770} & \multicolumn{1}{c|}{0.210} & \multicolumn{1}{c|}{0.733} & \multicolumn{1}{c|}{0.593} & \multicolumn{1}{c|}{0.793} & \multicolumn{1}{c|}{0.121} & \multicolumn{1}{c|}{0.664} & \multicolumn{1}{c|}{0.380} & \multicolumn{1}{c|}{0.689} & \multicolumn{1}{c|}{0.684} & \multicolumn{1}{c|}{0.137} & \multicolumn{1}{c|}{0.792} & \multicolumn{1}{c|}{0.576} & \multicolumn{1}{c|}{0.917} \\ \hline
CNF$^{\blacktriangle att}_{m4a}$ & \multicolumn{1}{c|}{0.856} & \multicolumn{1}{c|}{0.350} & \multicolumn{1}{c|}{0.747} & \multicolumn{1}{c|}{0.186} & \multicolumn{1}{c|}{0.844} & \multicolumn{1}{c|}{0.434} & \multicolumn{1}{c|}{0.766} & \multicolumn{1}{c|}{0.104} & \multicolumn{1}{c|}{0.558} & \multicolumn{1}{c|}{0.086} & \multicolumn{1}{c|}{0.648} & \multicolumn{1}{c|}{0.662} & \multicolumn{1}{c|}{0.123} & \multicolumn{1}{c|}{0.802} & \multicolumn{1}{c|}{0.478} & \multicolumn{1}{c|}{0.956} \\ \hline
CNF$^{\blacktriangle sum}_{m4a}$ & \multicolumn{1}{c|}{0.857} & \multicolumn{1}{c|}{0.352} & \multicolumn{1}{c|}{0.759} & \multicolumn{1}{c|}{0.199} & \multicolumn{1}{c|}{0.816} & \multicolumn{1}{c|}{0.448} & \multicolumn{1}{c|}{0.763} & \multicolumn{1}{c|}{0.105} & \multicolumn{1}{c|}{0.619} & \multicolumn{1}{c|}{0.248} & \multicolumn{1}{c|}{0.711} & \multicolumn{1}{c|}{0.678} & \multicolumn{1}{c|}{0.136} & \multicolumn{1}{c|}{0.826} & \multicolumn{1}{c|}{0.470} & \multicolumn{1}{c|}{0.958} \\ \hline
BRN$^{\blacktriangle att}_{fma}$ & \multicolumn{1}{c|}{0.881} & \multicolumn{1}{c|}{\underline{0.431}} & \multicolumn{1}{c|}{0.764} & \multicolumn{1}{c|}{0.210} & \multicolumn{1}{c|}{0.672} & \multicolumn{1}{c|}{0.580} & \multicolumn{1}{c|}{0.780} & \multicolumn{1}{c|}{0.116} & \multicolumn{1}{c|}{0.620} & \multicolumn{1}{c|}{0.312} & \multicolumn{1}{c|}{0.709} & \multicolumn{1}{c|}{0.684} & \multicolumn{1}{c|}{0.135} & \multicolumn{1}{c|}{0.837} & \multicolumn{1}{c|}{0.588} & \multicolumn{1}{c|}{0.942} \\ \hline
BRN$^{\blacktriangle att}_{m4a}$ & \multicolumn{1}{c|}{0.862} & \multicolumn{1}{c|}{0.367} & \multicolumn{1}{c|}{0.766} & \multicolumn{1}{c|}{0.208} & \multicolumn{1}{c|}{\underline{0.887}} & \multicolumn{1}{c|}{0.441} & \multicolumn{1}{c|}{0.788} & \multicolumn{1}{c|}{0.119} & \multicolumn{1}{c|}{0.610} & \multicolumn{1}{c|}{0.354} & \multicolumn{1}{c|}{0.706} & \multicolumn{1}{c|}{0.684} & \multicolumn{1}{c|}{0.144} & \multicolumn{1}{c|}{0.893} & \multicolumn{1}{c|}{0.574} & \multicolumn{1}{c|}{\underline{0.983}} \\ \hline
BRN$^{\blacktriangle sum}_{m4a}$ & \multicolumn{1}{c|}{0.870} & \multicolumn{1}{c|}{0.383} & \multicolumn{1}{c|}{0.767} & \multicolumn{1}{c|}{0.212} & \multicolumn{1}{c|}{0.836} & \multicolumn{1}{c|}{0.586} & \multicolumn{1}{c|}{0.793} & \multicolumn{1}{c|}{0.121} & \multicolumn{1}{c|}{0.629} & \multicolumn{1}{c|}{0.321} & \multicolumn{1}{c|}{0.740} & \multicolumn{1}{c|}{0.688} & \multicolumn{1}{c|}{0.140} & \multicolumn{1}{c|}{0.878} & \multicolumn{1}{c|}{0.583} & \multicolumn{1}{c|}{0.978} \\ \hline
\hline
CNF$^{\triangle att}$ & \multicolumn{1}{c|}{0.881} & \multicolumn{1}{c|}{0.422} & \multicolumn{1}{c|}{0.755} & \multicolumn{1}{c|}{0.201} & \multicolumn{1}{c|}{0.769} & \multicolumn{1}{c|}{0.510} & \multicolumn{1}{c|}{0.774} & \multicolumn{1}{c|}{0.109} & \multicolumn{1}{c|}{0.588} & \multicolumn{1}{c|}{0.254} & \multicolumn{1}{c|}{0.686} & \multicolumn{1}{c|}{0.677} & \multicolumn{1}{c|}{0.137} & \multicolumn{1}{c|}{0.814} & \multicolumn{1}{c|}{0.590} & \multicolumn{1}{c|}{0.953} \\ \hline
CNF$^{\blacktriangle att}$ & \multicolumn{1}{c|}{0.877} & \multicolumn{1}{c|}{0.420} & \multicolumn{1}{c|}{0.760} & \multicolumn{1}{c|}{0.200} & \multicolumn{1}{c|}{0.714} & \multicolumn{1}{c|}{0.572} & \multicolumn{1}{c|}{0.777} & \multicolumn{1}{c|}{0.110} & \multicolumn{1}{c|}{0.609} & \multicolumn{1}{c|}{0.290} & \multicolumn{1}{c|}{0.679} & \multicolumn{1}{c|}{0.672} & \multicolumn{1}{c|}{0.131} & \multicolumn{1}{c|}{0.787} & \multicolumn{1}{c|}{0.518} & \multicolumn{1}{c|}{0.906} \\ \hline
BRN$^{\triangle att}$ & \multicolumn{1}{c|}{0.880} & \multicolumn{1}{c|}{0.426} & \multicolumn{1}{c|}{0.751} & \multicolumn{1}{c|}{0.201} & \multicolumn{1}{c|}{0.654} & \multicolumn{1}{c|}{0.483} & \multicolumn{1}{c|}{0.769} & \multicolumn{1}{c|}{0.107} & \multicolumn{1}{c|}{0.666} & \multicolumn{1}{c|}{0.282} & \multicolumn{1}{c|}{0.675} & \multicolumn{1}{c|}{0.679} & \multicolumn{1}{c|}{0.134} & \multicolumn{1}{c|}{0.852} & \multicolumn{1}{c|}{0.590} & \multicolumn{1}{c|}{0.950} \\ \hline
BRN$^{\blacktriangle att}$ & \multicolumn{1}{c|}{\underline{0.884}} & \multicolumn{1}{c|}{0.414} & \multicolumn{1}{c|}{\underline{0.777}} & \multicolumn{1}{c|}{0.222} & \multicolumn{1}{c|}{\textbf{0.890}} & \multicolumn{1}{c|}{0.586} & \multicolumn{1}{c|}{0.802} & \multicolumn{1}{c|}{0.125} & \multicolumn{1}{c|}{\underline{0.687}} & \multicolumn{1}{c|}{0.335} & \multicolumn{1}{c|}{\textbf{0.759}} & \multicolumn{1}{c|}{0.695} & \multicolumn{1}{c|}{0.143} & \multicolumn{1}{c|}{\underline{0.900}} & \multicolumn{1}{c|}{0.610} & \multicolumn{1}{c|}{0.980} \\ \hline
BRN$^{\blacktriangle sum}$ & \multicolumn{1}{c|}{0.870} & \multicolumn{1}{c|}{0.397} & \multicolumn{1}{c|}{0.776} & \multicolumn{1}{c|}{\underline{0.227}} & \multicolumn{1}{c|}{0.845} & \multicolumn{1}{c|}{\underline{0.666}} & \multicolumn{1}{c|}{\underline{0.803}} & \multicolumn{1}{c|}{\underline{0.130}} & \multicolumn{1}{c|}{0.666} & \multicolumn{1}{c|}{\textbf{0.448}} & \multicolumn{1}{c|}{\underline{0.756}} & \multicolumn{1}{c|}{\underline{0.708}} & \multicolumn{1}{c|}{\underline{0.155}} & \multicolumn{1}{c|}{0.884} & \multicolumn{1}{c|}{\textbf{0.647}} & \multicolumn{1}{c|}{\textbf{0.992}} \\ \hline
\hline
MERT-$330$ & \multicolumn{1}{c|}{\textbf{0.891}} & \multicolumn{1}{c|}{\textbf{0.432}} & \multicolumn{1}{c|}{\textbf{0.786}} & \multicolumn{1}{c|}{\textbf{0.235}} & \multicolumn{1}{c|}{0.856} & \multicolumn{1}{c|}{\textbf{0.689}} & \multicolumn{1}{c|}{\textbf{0.816}} & \multicolumn{1}{c|}{\textbf{0.137}} & \multicolumn{1}{c|}{\textbf{0.714}} & \multicolumn{1}{c|}{\underline{0.413}} & \multicolumn{1}{c|}{0.659} & \multicolumn{1}{c|}{\textbf{0.717}} & \multicolumn{1}{c|}{\textbf{0.163}} & \multicolumn{1}{c|}{\textbf{0.928}} & \multicolumn{1}{c|}{\underline{0.642}} & \multicolumn{1}{c|}{0.905} \\ \hline

\end{tabular}%
}
\caption{Experiment results of our various models on the downstream task framework. We build upon our previous notation by incorporating the trained dataset as a subscript to each model. This dataset subscript is specifically added only for the public datasets we used in our experiments. For downstream tasks that involve multiple datasets, each dataset is listed separately, along with its corresponding evaluation metrics.}
\label{table:results}
\end{table*}

In this section, we discuss the results of our experiments. The performance metrics for the evaluation of downstream tasks are presented in Table \ref{table:results}. We evaluate our models from three different perspectives: the effect of the dataset size, the number of model parameters, and the substitution of the multi-head self-attention module with SummaryMixing.

During experimentation, we further examined both of our backbone models using their pretrained weights on speech data. We then performed the downstream task training and evaluation as described earlier. The performance of those models was much lower compared to the rest of our setup and for this reason we omit their results from the discussion.

\subsection{Effects of dataset size}
First of all, it is important to note that all models were trained for the same number of steps. This consistency in training duration means that the models iterated over the small public datasets significantly more times compared to our large proprietary dataset. Specifically, since the smaller datasets contain fewer examples, the models were exposed to them repeatedly in each training cycle, resulting in a higher number of passes through the data. In contrast, the larger proprietary dataset, with its larger volume of data, led to fewer iterations over each individual example during the same number of training steps. This difference in the frequency of exposure to the data could potentially influence the models' performance and generalization capabilities, as smaller datasets may allow for more frequent learning updates per data point.

Comparing models BRN$^{\blacktriangle att}$, BRN$^{\blacktriangle att}_{fma}$ and BRN$^{\blacktriangle att}_{m4a}$, we observe that they achieve similar performance even when trained on significantly smaller datasets. Our proprietary dataset contains around $200k$ hours of music while FMA and M4A just under $1k$ hours. On the majority of the metrics the difference between the large and small datasets is not more than a couple of percentage points. However, this becomes more noticeable when looking into the tasks of emotion score regression, instrument classification, pitch classification, vocal technique detection and singer classification. The largest differences can be observed in key detection for FMA ($-21.8\%$) and genre classification (GTZAN) for M4A ($-14.5\%$). 

The same phenomenon is observed in the Conformer models. CNF$^{\blacktriangle att}_{fma}$ is, in fact, performing slightly better than CNF$^{\blacktriangle att}$, with a small increase in most metrics, with emotion score regression and vocal technique detection being standouts. In contrast, CNF$^{\blacktriangle att}_{m4a}$ performs worse than the other two in most tasks, with large drops in genre classification ($-15.9\%$ GTZAN), emotion score regression ($-10.6\%$ $r^2_A$ / $-29.4\%$ $r^2_V$) and vocal technique detection ($-9.8\%$). Surprisingly, it performs better in key detection ($+11.1\%$). Comparing only our large proprietary dataset with M4A, it would be a fair assumption to attribute these drops just to the size difference between the datasets. Taking into account the comparison between the similar sized datasets, however, we still observe that M4A performs worse. This leads us to believe that this is probably in part due to the nature of the audio content, which favors some tasks over others.

\subsection{Architecture differences}
In the context of model size, we first compare models with the same base architecture, where the reduction in parameters is due to the lower number of layers. The larger Branchformer model BRN$^{\blacktriangle att}$ generally outperforms the smaller BRN$^{\triangle att}$, with the largest increases being in key detection ($+23.6\%$), genre classification ($+10.3\%$ GTZAN) and instrument classification ($+8.4\%$ nsynth). The increase in all the other tasks is lower but noticeable. Comparing the Conformer models CNF$^{\blacktriangle att}$ and CNF$^{\triangle att}$, we see that they perform similarly, with no model being strictly better in the majority of tasks. However, CNF$^{\blacktriangle att}$ performs noticeably better in genre classification (GTZAN), while CNF$^{\triangle att}$ performs better in vocal technique detection and singer identification. In both cases the smaller model demonstrate competitive results with only around $30\%$ of the parameters. This observation shows that the performance of a model does not always scale with the number of parameters and that smaller models have the potential to compete.

We also evaluate the case where we have less parameters due to different base architectures, all other things being the same. Here, BRN$^{\blacktriangle att}$ outperforms CNF$^{\blacktriangle att}$, having only $57\%$ of its parameters. Emotion score regression, instrument classification and singer identification are some standouts but the greatest gains are in key detection ($+17.6\%$), pitch classification ($+11.3\%$) and vocal technique detection ($+9.2\%$). In the other tasks the increase in performance is lower. We additionally observe that the smaller versions have similar results. In this case, BRN$^{\triangle att}$ has $63\%$ of the parameters of CNF$^{\triangle att}$, but the relative metric difference between them in most tasks is very low. Notable exceptions are: key detection in favor of CNF$^{\triangle att}$ and emotion score regression in favor of BRN$^{\triangle att}$. These results present another indication that the capabilities of smaller models can originate not only from a plain decrease in parameters, but also from a different architectural configuration.

\subsection{SummaryMixing vs Multi-head self-attention}
The substitution of multi-head self-attention with SummaryMixing also leads to similar, if not better, performance. With $8.5\%$ less parameters, CNF$^{\blacktriangle sum}_{m4a}$ consistently performs slightly better than CNF$^{\blacktriangle att}_{m4a}$, with a small increase in the metrics of most downstream tasks. This increase is larger when looking into the tasks of emotion score regression ($+6.1\%$ $r^2_A$ / $+16.2\%$ $r^2_V$) and instrument classification ($+6.3\%$ Nsynth). 

The decrease in parameters goes up to $12.3\%$ in the case of the Branchformer models. Again, regardless of whether we trained on M4A or our proprietary dataset, the relative difference between most metrics is relatively small. However, the SummaryMixing module performs better on genre classification (GTZAN) on both datasets ($+14.5\%$ M4A / $+8\%$ proprietary) and on emotion score regression on the proprietary dataset only ($+11.3\%$ $r^2_V$ only). The attention module performs better in key detection on both datasets.

As discussed above, the SummaryMixing module shows potential and benefits both the Conformer and Branchformer models. There is a non-negligible decrease in model parameters while maintaining competitive performance. Our findings are also consistent with those of the original work~\cite{parcolletsummarymixing}, showing the potential of SummaryMixing when applied to the field of music information retrieval.

\subsection{Comparison with state-of-the-art}
Our two best performing models over all downstream tasks are BRN$^{\blacktriangle att}$ and BRN$^{\blacktriangle sum}$. When compared with the current state-of-the-art MERT-$330$ model, we see that our model is very competitive and achieves very similar results in the majority of tasks. MERT performs only slightly better overall against both our models, with the only notable exceptions being vs BRN$^{\blacktriangle att}$ on genre classification (GTZAN) ($+10.3\%$) and emotion score regression ($+7.8\%$ $r^2_V$ only). However, the difference is reduced vs BRN$^{\blacktriangle sum}$. On genre classification (GTZAN), MERT is only better by a smaller margin and on emotion score regression it even performs slightly worse. Furthermore, both our models achieve better results in instrument classification (nsynth) ($+10\%$ BRN$^{\blacktriangle att}$ / $+9.7\%$ BRN$^{\blacktriangle sum}$) and singer identification ($+7.5\%$ BRN$^{\blacktriangle att}$ / $+8.7\%$ BRN$^{\blacktriangle sum}$). We should note that a potential explanation for these larger differences might lie in the composition of the large proprietary datasets on which the models where trained. Still, our proposed architecture managed to achieve results very close to the state-of-the-art, while reducing model size.

\section{Discussion \& Future Work}\label{sec:discussion}

Although our downstream evaluation covers a diverse set of tasks, additional extensions to include challenging domains such as audio matching, cover song recognition, and hit song detection would provide a more comprehensive assessment of foundation model capabilities. Due to limited computational resources, we were unable to perform extensive hyperparameter tuning, including alternative masking strategies, tokenization methods, different embedding layer selection, and cross-domain training, e.g. integration of music and speech datasets, or starting from speech pretrained weights, instead of random initialization. Additionally, the current model processes $30$-second audio segments with fixed $400$ ms subsegments. We aim to exploring longer contexts, finer granularity, or overlapping segments could further enhance representational quality and downstream performance.
We also intend to compare common model compression techniques, including pruning, quantization, and distillation, to evaluate their performance in reducing model size.

\section{Conclusion}\label{sec:conclusions}

The goal of this work is to investigate the ability of an audio foundation model to maintain competitive performance on a variety of downstream MIR tasks, while reducing its size. We use the Branchformer architecture in the field of music, which was previously used in speech recognition. Similarly, we combine it with a promising alternative to multi-head self-attention, namely SummaryMixing. 
We conducted pre-training using publicly available datasets, along with a proprietary dataset.
We showed that these approaches have the potential to achieve competitive results in many downstream tasks and at the same time reduce the number of model hyperparameters anywhere from $8.5$\% to $12.3$\%, when trained on a comparably large corpus of music data. 
Furthermore, we achieve state-of-the-art results in some downstream tasks, such as key detection, instrument classification, and singer identification, improving previous metrics by up to $10$\%. Our contribution further cements the abilities and flexibility of foundation models in music information retrieval.


\printbibliography
 
\appendix

\end{document}